\documentclass[twoside,letterpaper]{pazh_mod}

\usepackage{textcomp,mathptmx}
\usepackage{graphicx}

\begin{document}

\journalinfo{2003}{29}{7}{425}[428]

\title{A Possible Shock Wave in the Intergalactic Medium of the Cluster of
  Galaxies A754} 

\author{R.A.\ Krivonos\email{krivonos@hea.iki.rssi.ru}\address{1}, 
   A.A.\ Vikhlinin\address{1,2}, 
   M.L.\ Markevitch\address{2}, 
   M.N.\ Pavlinsky\address{1}
   \addresstext{1}{Space Research Institute, Russian Academy of Sciences, Profsoyuznaya ul. 84/32, Moscow 117810, Russia}
   \addresstext{2}{Harvard-Smithsonian Center for Astrophysics, 60 Garden Street, Cambridge MA 02138, USA} 
}

\shorttitle{POSSIBLE SHOCK WAVE IN A754} \shortauthor{KRIVONOS ET AL.} 
\submitted{Accepted for publication in Astron.Lett.}

\begin{abstract}
  The cluster of galaxies A754 undergoes a merger of several large
  structural units. X-ray observations show a nonequilibrium state of the
  central part of the cluster, in which a cloud of cold plasma 500 kpc in
  size was identified amid the hotter cluster gas. The X-ray image of A754
  exhibits a brightness discontinuity, which can be interpreted as a shock
  wave in front of a moving cloud of dense gas.The shock parameters are
  determined from the jump in intergalactic gas density using the ROSAT
  image. The estimated Mach number is $M_1=1.71^{+0.45}_{-0.24}$ at a 68\%
  confidence level.

\end{abstract}

\section{INTRODUCTION}

Clusters of galaxies are the largest gravitationally bound objects in the
Universe. Mergers of clusters result in the release of potential energy into
the intergalactic medium at a level of $10^{63}$ erg, mainly in the form of
heating by shock waves. Therefore, observations of shock waves in the
intergalactic medium are particularly interesting because they provide
information for studying the physical processes that accompany the cluster
formation.

In this paper we study A754, a rich cluster of galaxies at z = 0.0541 in the
stage of violent formation. ASCA observations of this cluster clearly
indicate that the central part of the cluster is in a nonequilibrium state
and hot and cold regions are identifiable in it (Henriksen and Markevitch
1996).The pattern probably represents the motion of a cold dense cloud
through a hotter ambient gas.  Structures of this kind were observed by the
Chandra satellite in several clusters (Markevitch et al. 2000; Vikhlinin et
al. 2001; Mazzota et al.  2001).The merging parts of the clusters are
generally expected to move at a supersonic speed, which leads to the
formation of shock waves.

The ROSAT X-ray image of A754 exhibits a surface brightness discontinuity,
which can be interpreted as a shock wave in front of a moving cloud of dense
gas. The amplitude of the surface brightness discontinuity makes it possible
to determine the jump in gas density at the shock front and, hence, the Mach
number $M=v/v_s$, where $v_s$ is the shock velocity and $v_s$ is the speed
of sound in the medium. Unfortunately, the ROSAT energy range (0.5-2.5 keV)
does not allow us to determine the gas temperature at the shock boundary;
i.e., we cannot independently verify the interpretation of the brightness
discontinuity as a shock wave using the temperature jump.

We quote the physical length scales assuming
$H_0=50$~km~s$^{-1}$~Mpc$^{-1}$. The derive shock velocity and Mach number
are independent of $H_0$.

\section{DATA ANALYSIS}

The ROSAT image of the cluster of galaxies A754 is shown in Figure 1. The
region of cold gas appears as a bright cloud at the image center. The shape
and location of a brightness discontinuity to the left from the center are
consistent with those of a shock in front of a cloud of cold gas moving at a
supersonic speed.

Quantitative information about the gas density distribution can be obtained
by analyzing the surface brightness profile. The shape of the front of the
putative shock is well described by a circle with a radius of 775 kpc. We
use the center of this circle as the reference point for radial
distances. The surface brightness profile was extracted in a $\pm15^{\circ}$
sector with respect to the front symmetry axis\footnote{The sector opening
angle was determined by the size of the brightness discontinuity region; the
brightness profile for the outer part of the shock was obtained in a wider
sector ($\pm20^{\circ}$) to improve statistics.}. The profile was measured
in concentric rings of equal logarithmic width; the ratio of the inner and
outer ring radii was 1.01. We verified that varying the reference point
changed the measured jump in gas density only slightly.

\begin{figure*}[htb]
  \centerline{\includegraphics[width=0.7\linewidth]{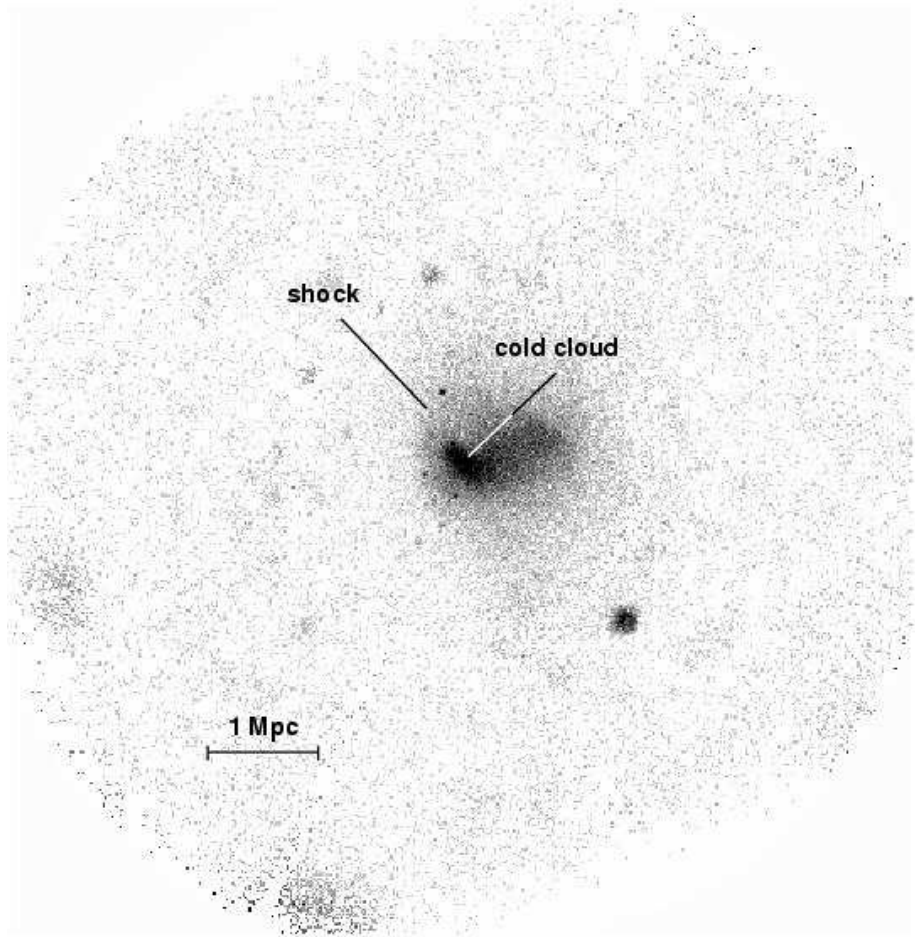}}
  \caption{
    The ROSAT X-ray image of A754 in the 0.7--2.0 keV energy band. The
    brightness discontinuity to the left from the central body most likely
    corresponds to the shock.}
  \label{fig:data}
\end{figure*}

Measuring the gas density distribution at large radii requires careful
background subtraction and corrections for the telescope vignetting. We used
S.~Snowden's software (Snowden et al. 1994) for the data reduction. The
software allows one to eliminate the periods of an anomalously high
background of particles and scattered solar X-ray emission. In addition, the
exposure maps are computed and all of the known background components are
fitted and subtracted from the image. The final result is the image in a
given energy range that contains only the cluster radiation, point-like X-ray
sources, and the cosmic X-ray background. An optimal ratio of the cluster
surface brightness to the background level is achieved by using data above
0.7 keV.

When measuring the X-ray brightness profile, we eliminated all of the
detectable point-like sources and all the extended sources unrelated to the
main cluster emission. The cluster makes a significant contribution to the
total brightness even at large distances from the center ($\sim2\textrm{
Mpc}$). Therefore, it is hard to identify a region that could be used to
directly determine the background level. Instead, we derived the background
level by fitting the brightness profile at large distances with the model
consisting of a $\beta$-model (Cavaliere and Fusco-Femiano 1976) representing
the cluster emission and a constant component representing background:
\begin{equation}
I(r)=I_0\left(1+\left(r/r_{c}\right)^2\right)^{-3\beta+1/2}+I_{b}
\end{equation}
where $I(r)$ is the surface brightness, $I_{b}$ is the background intensity,
and $r_{c}$ is the core radius.

\section{FITTING THE BRIGHTNESS PROFILE}

The main parameter of the shock motion is the Mach number, $M=v/v_s$.The
Mach number can be expressed in a standard way (Landau and Lifshitz 1988) in
terms of the jump in any of the gas parameters (density, pressure,
temperature) at the shock front. The most easily measurable quantity is the
density jump, because the gas density is related to the emissivity by a
simple relation: $\varepsilon\sim\rho^2$. Consequently, the problem reduces
to deprojecting the brightness discontinuity.

The brightness profile can be best deprojected by fitting the data by some
analytic dependence. Since the outer part of the cluster is most likely to
have not yet been perturbed by the shock, we can assume that the gas density
profile is described by the standard $\beta$-model. Inside the shock front,
we will be concerned with a narrow range of radii in which the density
profile can be assumed, to sufficient accuracy, to be a power-law function
of radius. Thus, we have the following model of the plasma emissivity:

\begin{equation}\label{model}
\varepsilon(r) = \left\{ \begin{array}{ll}
 \varepsilon_{2}\left(r/R\right)^{\alpha} & \textrm{если $r\le R$}\\
 \varepsilon_{\beta}\left(1+r^{2}/r_{c}^{2}\right)^{-3\beta} & \textrm{если $r>R$}\\
  \end{array} \right.
\end{equation}

where R is the front position, $\varepsilon_{2}$ is the post-shock
emissivity, and $\varepsilon_{\beta}$ is the central emissivity for the
$\beta$ model. The core radius $r_{c}$ was fixed at a typical value of 250
kpc (Jones and Forman 1984).

\begin{figure*}
        \includegraphics[width=0.5\textwidth]{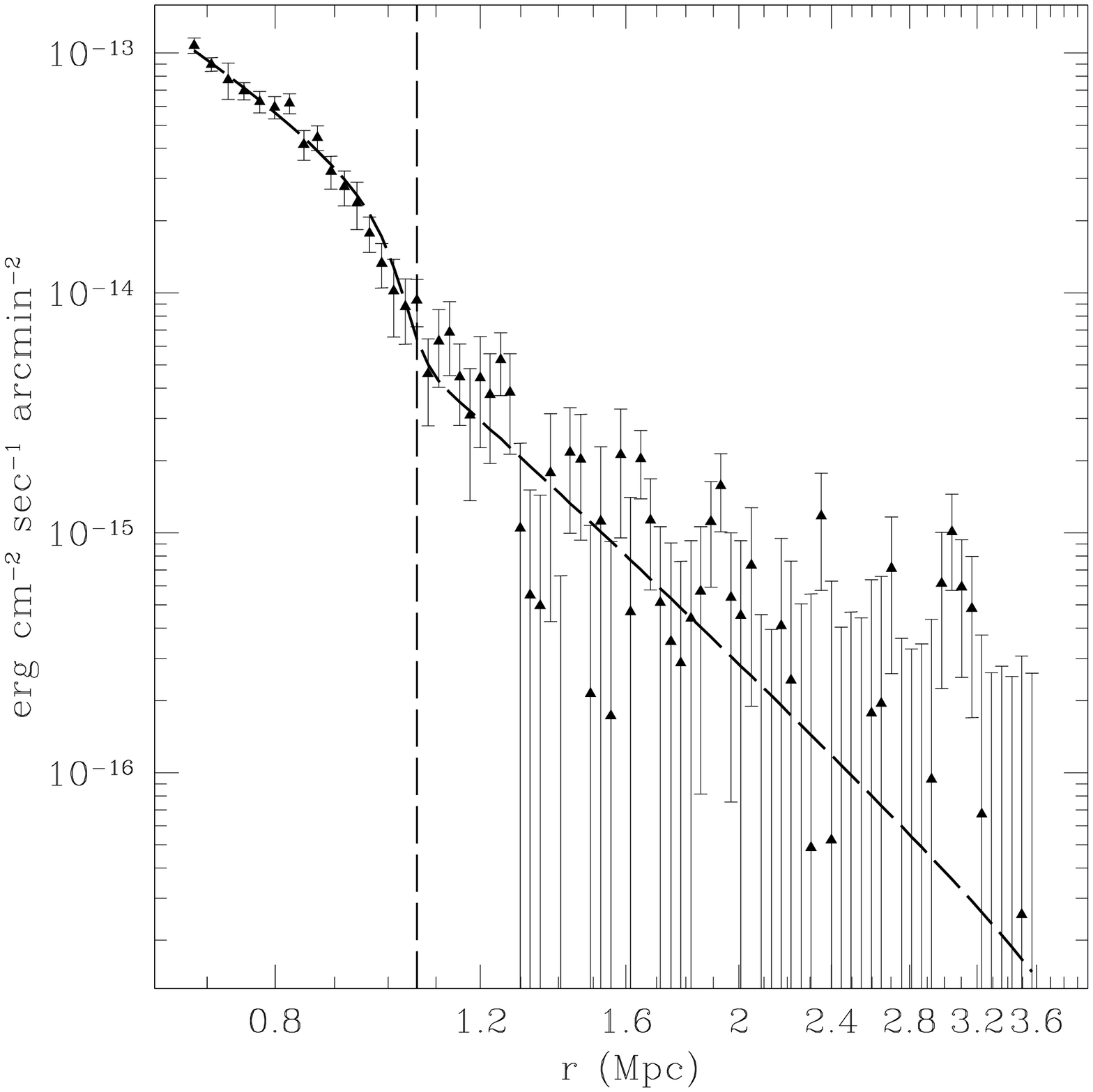}
        \includegraphics[width=0.5\textwidth]{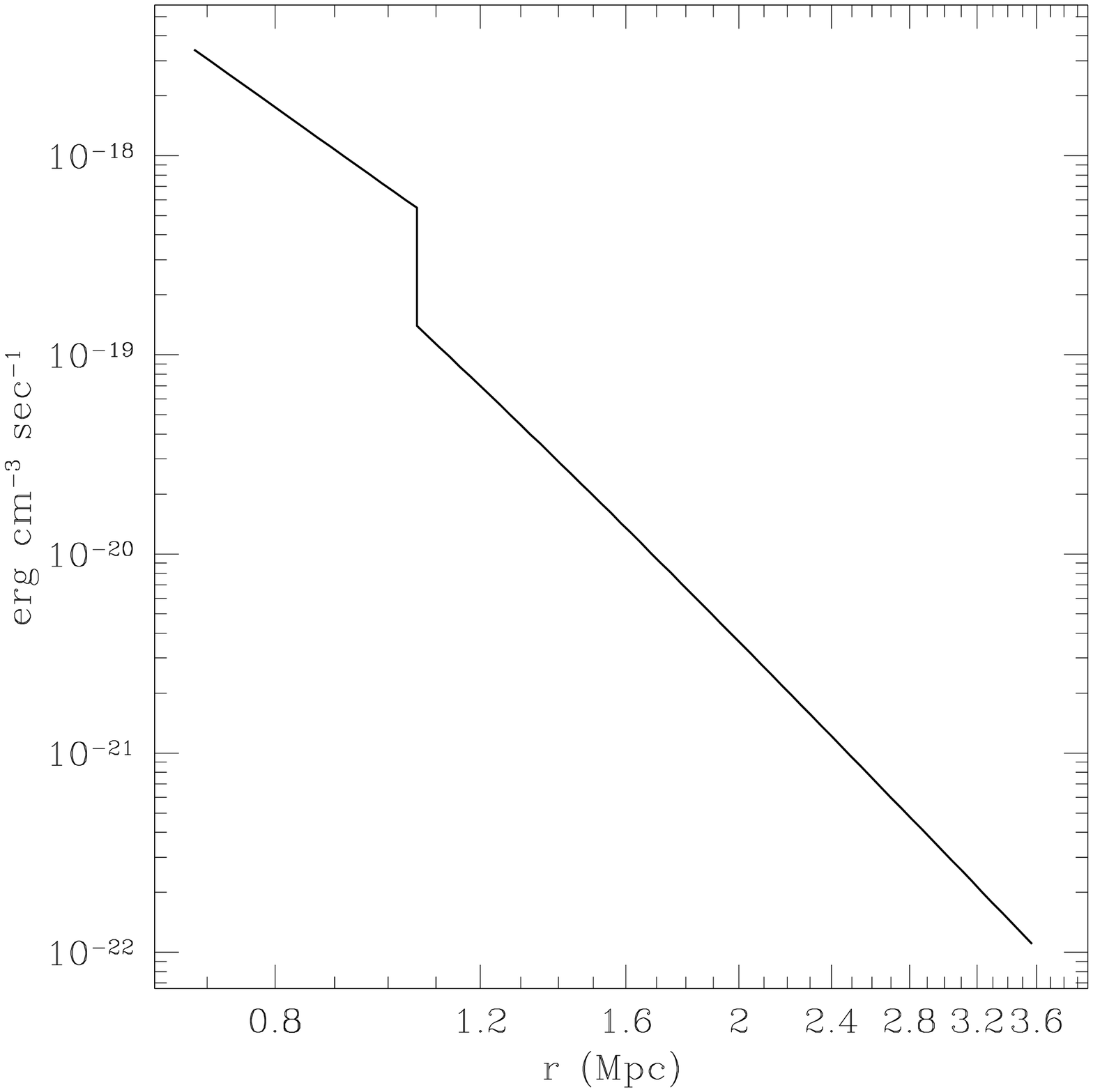}
        \caption{
          (a) The surface brightness profile in the shock region. The
          distances are measured from the center of curvature of the front
          along the shock propagation; the solid line represents the fitting
          curve obtained by integrating the plasma emissivity model (b)
          along the line of sight.
        }
\label{fig:prof}
\end{figure*}

The model emissivity profile was numerically integrated along the line of
sight and folded with the angular resolution of the telescope
($\sim25''$). Note that the normalization of the $\beta$ model in
equation~(\ref{model}) can be expressed in terms of the pre-shock emissivity,
$\varepsilon_1$.  To determine the parameters of the above model, we
minimized the $\chi^{2}$ value for four parameters
($\varepsilon_{2},\varepsilon_{2}/\varepsilon_{1},\beta,\alpha$). Figure
\ref{fig:prof} shows the derived model of the gas emissivity distribution
(b) and the surface brightness profile (a).

\section{RESULTS}

\begin{figure*}
  \includegraphics[width=0.5\textwidth]{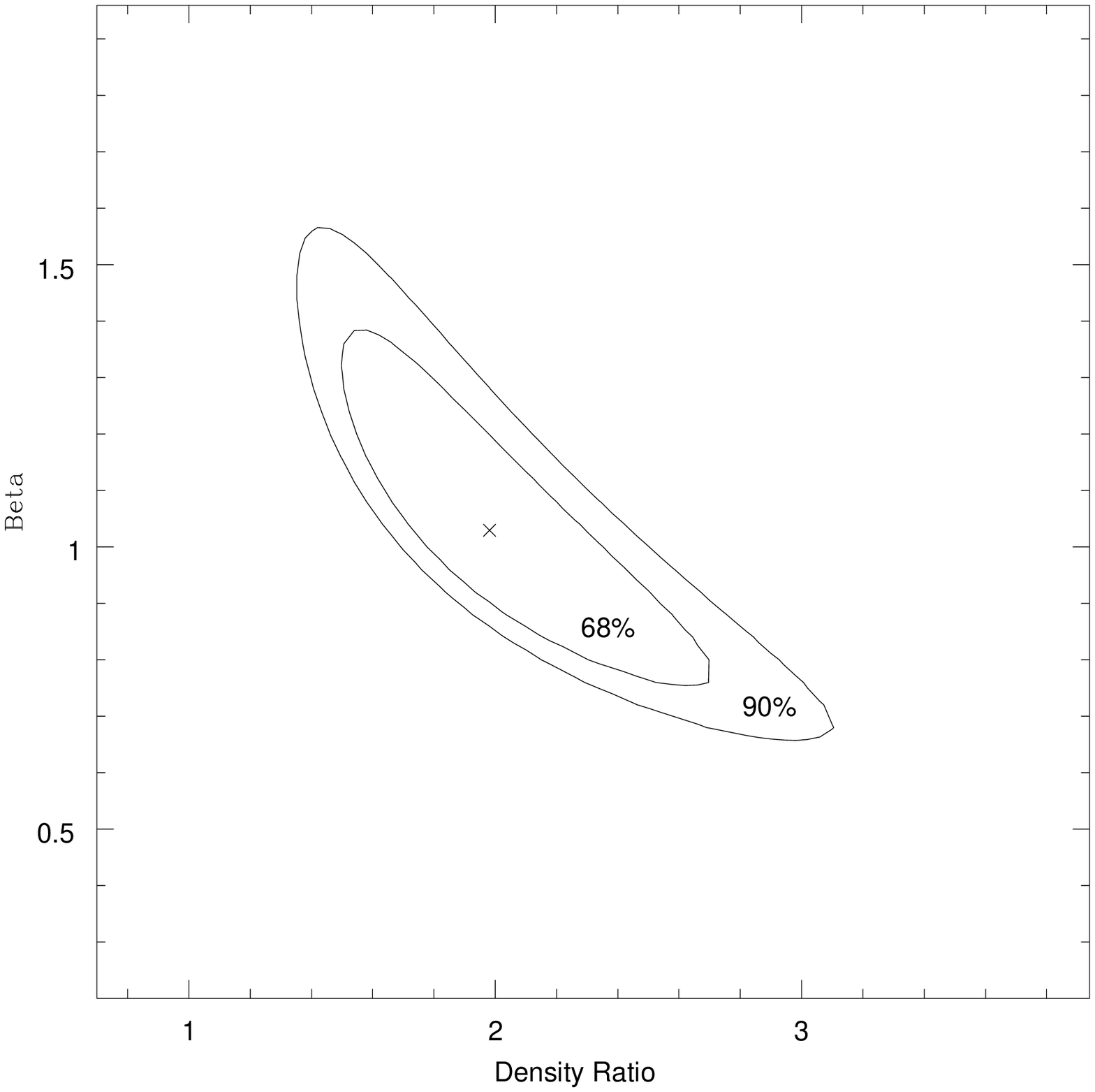}
  \includegraphics[width=0.5\textwidth]{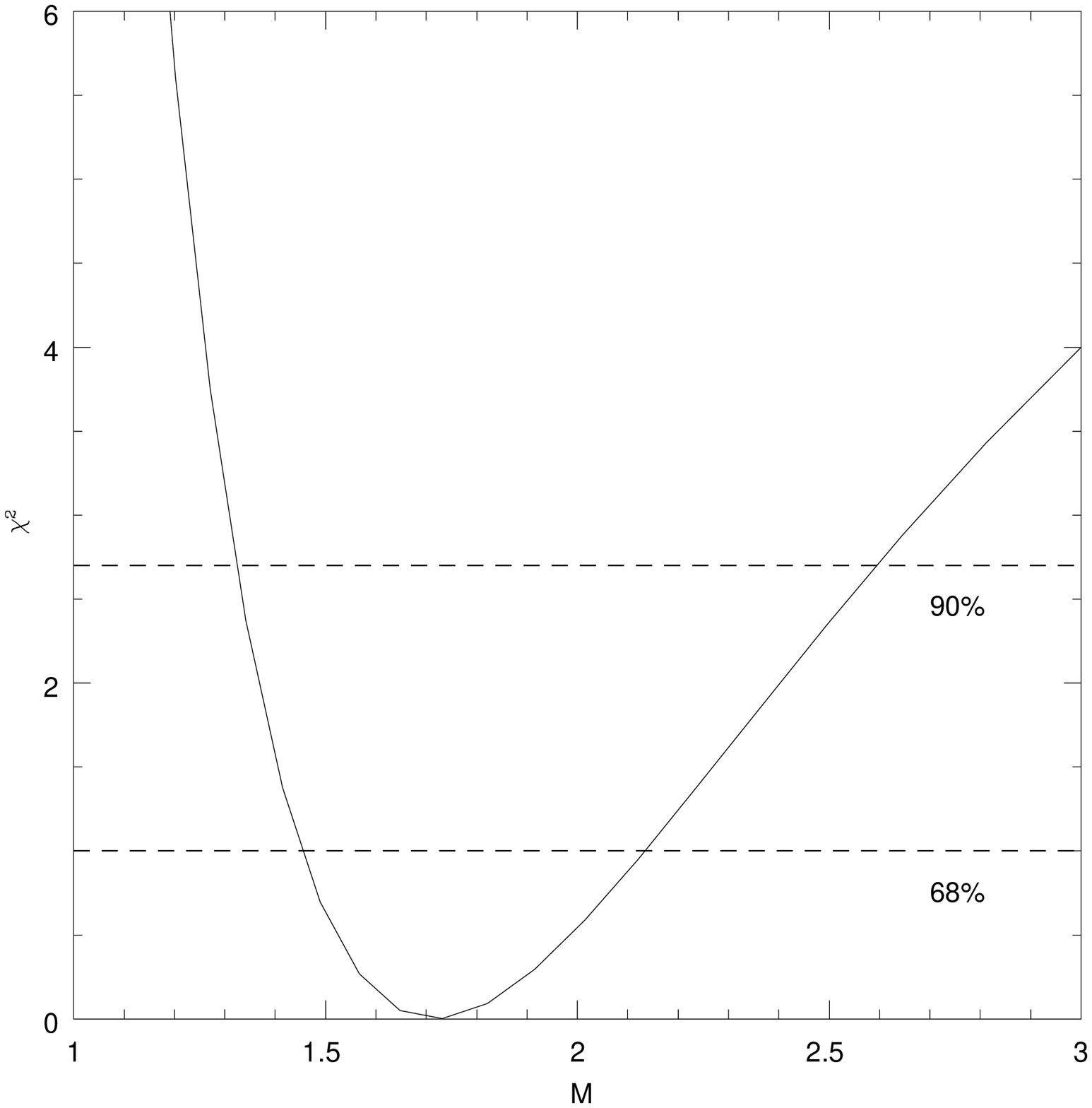}
    \caption{
      The 68\% and 90\% confidence regions for two parameters of our model
      of the intergalactic gas radiation: the gas density ratio at the shock
      boundary and the parameter $\beta$ in the outer part of the cluster
      [see Eq. (\ref{model})]. (a) The cross marks the point that
      corresponds to the best-fit parameters; (b) $\chi^2$ versus Mach
      number.
    }
   \label{fig:data1}
\end{figure*}

The emissivity jump obtained by fitting the brightness profile is trivially
transformed to the density jump ($\varepsilon_2/\varepsilon_1$):
\begin{equation}
 \frac{\rho_{2}}{\rho_{1}}=\left(\frac{\varepsilon_2}{\varepsilon_1}\right)^{1/2}=1.98^{+0.44}_{-0.32},
 \end{equation} 
where the uncertainties are at a 68\% confidence level for one interesting
parameter. From the $\rho_2/\rho_1$ ratio, we find the Mach number of the
shock with respect to the ambient gas:
\begin{equation}\label{eq:rho:M}
  \frac{\rho_{2}}{\rho_{1}}=\frac{(1+\gamma)\,M^2_{1}}{2+(\gamma-1)\,M^2_{1}}
\end{equation}
where $\gamma=5/3$ is the adiabatic index for monatomic gas. From
(\ref{eq:rho:M}), we obtain $M_{1}=1.71^{+0.45}_{-0.24}$ at a 68\% 1 =
1.71+0.45 -0.24 confidence level 68\% (see Fig.\ref{fig:data1}).

The derived value of $M_1$ corresponds to the gas temperature jump
$T_{2}/T_{1}=1.72^{+0.55}_{-0.28}$. In principle, measuring such a
temperature discontinuity would be an independent test of the validity of
the interpretation of the observed structure as a shock wave. Unfortunately,
the available Chandra and XMM observations of A754 do not allow us to
determine the gas temperature outside the shock because of the low
signal-to-noise ratio. Inside the shock, the temperature is measured
reliably, $T_2=10$~keV (Markevitch et al. 2003) from Chandra data.

A useful quantity is the shock velocity relative to the gas on the inner
side of the discontinuity:
\begin{equation}
M_2=\left(\frac{2+(\gamma-1)M_1^2}{2\gamma\,M_1^2-(\gamma-1)}\right)^{1/2}=0.66\pm0.07
\end{equation}
The measured value of $T_2$ allows us to determine the speed of sound and,
hence, the absolute shock velocity: $v_2=1070\pm115$~km~s$^{-1}$,
$v_1=2100_{-150}^{+200}$~km~$s^{-1}$.

The derived Mach number of a likely shock in A754 is close to the
theoretically expected values for merging clusters: $M=2-3$ (Sarazin
2002). Comparison with other clusters is hard because observations of merger
shocks in clusters are is still rare. In two other best-studied cases,
(1E0657-56--Markevitch et al.{} 2002, Cyg A--Markevitch et al.{} 1999), high
Mach numbers are also inferred. However, in A3667 one observes a merger with
$M\sim1$ (Vikhlinin et al.{} 2001). Clearly, detailed studies of a larger
sample of merging clusters would be valuable.

\section{ACKNOWLEDGMENTS}

This work was supported by the Russian Basic Research Foundation (grants
00-02-1724 and 00-15-96699), and by the ``Young Scientist'' and
``Nonstationary Astronomical Phenomena'' programs of the Russian Academy of Sciences.

\end{document}